**Title:** Nanoscale trapping of interlayer excitons in a 2D semiconductor heterostructure

**Author Names:** Daniel N. Shanks[1], Fateme Mahdikhanysarvejahany[1], Christine Muccianti[1], Adam Alfrey[1], Michael R. Koehler[2], David G. Mandrus[3-5], Takashi Taniguchi[6], Kenji Watanabe[7], Hongyi Yu[8], Brian J. LeRoy[1], and John R. Schaibley[1]*

**Author Addresses:**
[1]Department of Physics, University of Arizona, Tucson, Arizona 85721, USA
[2]JIAM Diffraction Facility, Joint Institute for Advanced Materials, University of Tennessee, Knoxville, TN 37920
[3]Department of Materials Science and Engineering, University of Tennessee, Knoxville, Tennessee 37996, USA
[4]Materials Science and Technology Division, Oak Ridge National Laboratory, Oak Ridge, Tennessee 37831, USA
[5]Department of Physics and Astronomy, University of Tennessee, Knoxville, Tennessee 37996, USA
[6]International Center for Materials Nanoarchitectonics, National Institute for Materials Science, 1-1 Namiki, Tsukuba 305-0044, Japan
[7]Research Center for Functional Materials, National Institute for Materials Science, 1-1 Namiki, Tsukuba 305-0044, Japan
[8]Guangdong Provincial Key Laboratory of Quantum Metrology and Sensing & School of Physics and Astronomy, Sun Yat-Sen University (Zhuhai Campus), Zhuhai 519082, China

**Corresponding Author:** John Schaibley, johnschaibley@email.arizona.edu

**Abstract:**

For quantum technologies based on single excitons and spins, the deterministic placement and control of a single exciton is a long-standing goal. $MoSe_2$-$WSe_2$ heterostructures host spatially indirect interlayer excitons (IXs) which exhibit highly tunable energies and unique spin-valley physics, making them promising candidates for quantum information processing. Previous IX trapping approaches involving moiré superlattices and nanopillars do not meet the quantum technology requirements of deterministic placement and energy tunability. Here, we use a nanopatterned graphene gate to create a sharply varying electric field in close proximity to a $MoSe_2$-$WSe_2$ heterostructure. The dipole interaction between the IX and the electric field creates a ~20 nm trap. The trapped IXs show the predicted electric field dependent energy, saturation at low excitation power, and increased lifetime, all signatures of strong spatial confinement. The demonstrated architecture is a crucial step towards deterministic trapping of single IXs, which has broad applications to scalable quantum technologies.

**Introduction:**

$MoSe_2$-$WSe_2$ heterostructures have a type-II band alignment, allowing for IXs that comprise an electron in the $MoSe_2$ layer Coulomb-bound to a hole in the $WSe_2$ layer.[1–15] Due to the spatially indirect nature of the IXs, they possess a permanent dipole moment which has been used to control their energy, lifetime, and transport.[2,16–18] Controlling the diffusion of IXs using a spatially varying electric field has already been established on a micron level scale,[17,18] and these efforts have shown that the valley degree of freedom can be conserved during transport.[2] These properties, in addition to the long lifetime of IXs,[1,19] and individual trapping capability, which has already been shown at moiré sites,[6,20] make them promising candidates for valleytronic quantum information applications.[4] However, these studies have yet to realize the deterministic placement and control of a singly trapped IXs, a basic building block for optically controlled quantum information systems in two-dimensional (2D) materials.[21]

In this letter, we demonstrate deterministic nanoscale confinement potentials that trap IXs, and investigate the low temperature optical response of these trapped IXs. Our approach is based on a 2D van der Waals heterostructure, with a nanopatterned gate to create a spatially varying electric field which traps IXs to a quantum dot-like potential, i.e., trapped IXs are spatially confined to the nanoscale in all three dimensions. Figure 1a shows a diagram of the device, which comprises a $MoSe_2$-$WSe_2$ heterostructure, fully encapsulated in hBN, with few layer graphene (FLG) top and bottom gates control the out-of-plane electric field applied to the heterostructure.[16,17] Figure S1 shows an optical microscope image of the structure. Device details including the exact layer thicknesses and transition metal dichalcogenide (TMD) relative twist angle are given in Methods.

**Results:**

We apply opposite polarity top ($V_{tg}$) and bottom ($V_{bg}$) gate voltages to produce an out-of-plane electric field without doping the TMD heterostructure.[16] The spatially varying electric field is constructed by etching a small hole in the FLG top gate, creating a nanoscale area under which the applied electric field is weaker. Atomic force microscopy topography of the patterned FLG shows that the hole diameters are roughly 30 nm (see Figure S2). Figure 1b shows results from a COMSOL simulation of the structure where the grey arrows depict the electric displacement amplitude and direction, clearly showing a reduction of the out-of-plane component under the hole.

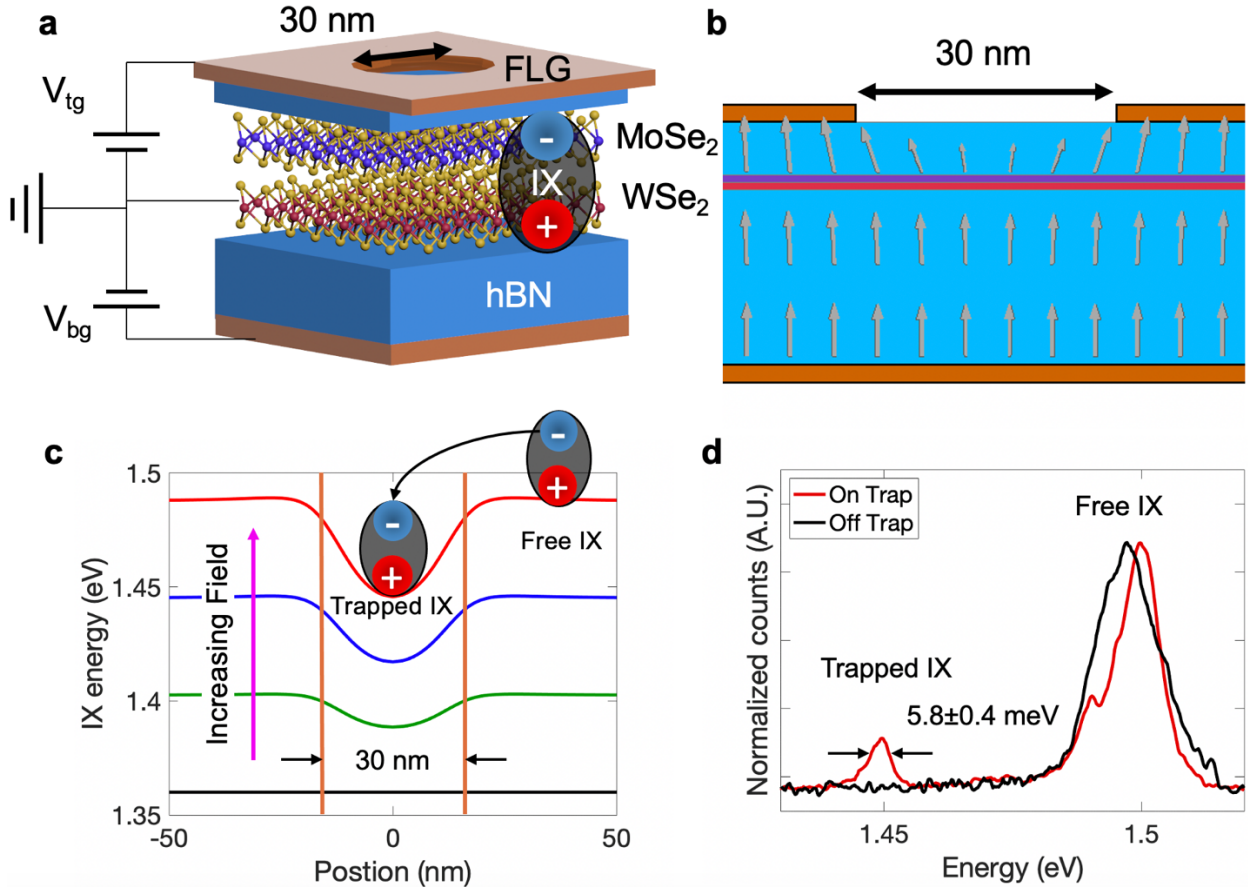

**Figure 1.** Nanoscale trapping of IXs in an MoSe$_2$-WSe$_2$ heterostructure. (a) Depiction of the nanopatterned heterostructure device. (b) COMSOL model, showing the profile of the electric field (grey arrows) with removed graphene. Graphene gates are shown in orange, hBN in blue, and TMD layers in purple and red. (c) Calculated IX energy as a function of position for four applied electric fields. Black line shows no applied field, green and blue show increasing field, and the red line shows IX potential energy profile for the maximum field applied to the device. Orange lines mark the edges of the graphene hole. (d) Confocal PL spectra of the free and trapped IX with collection points on (red) and off (black) of the nanoscale trap.

In our structure, the MoSe$_2$ (WSe$_2$) is on the top (bottom) so that the permanent dipole moment of the IX points down. With the electric field applied in the opposite direction of the permanent dipole moment of the IX, the energy of the IX is increased from its unperturbed energy due to the dipole energy shift, by H = −**p**·**E**, where H is the IX energy shift, **p** is the dipole moment of the IX, and **E** is the electric field. Because the field is weaker under the graphene hole, a spatial potential energy trap is created, where the IX energy shift is reduced proportionally to the strength of the field. Figure 1c depicts trapping of IXs using the COMSOL modelled electric field profile and the measured IX dipole moment for four applied electric fields. This model leads to a maximum trapping potential with a FWHM of 22 nm and depth of 42 meV. Fitting the maximum induced potential to a quantum harmonic oscillator solution leads to a ground state energy of 2.5 meV, and a ground state wavefunction with a spatial width of 6.1 nm, which is larger than both the IX Bohr radius of ~1-2 nm,[22] and the expected moiré lattice periodicity of 3.5 nm for the 54.5° twist angle of this device. Figure S3 shows the IX envelope wavefunction width as a function of applied electric field. While the PL signal was too weak to perform single photon emission experiments, we note that the strong dipolar repulsion between IXs prevents multiple occupancy of the same trap, and no signatures of trapped biexcitons were observed.[23] Figure 1d shows confocal photoluminescence (PL) spectra of the structure, with the gates set to maximum trapping potential configuration, for a collection spot on and off of the IX trap. With the collection spot off the trap, there is only a signal from the free IX at 1.497 eV. When the collection spot is aligned to the trap, an additional lower energy PL signal appears at 1.449 eV. Trapped IX signals from multiple hole sites and a secondary device are shown in Figures S4-5. This result confirms the controlled placement of the IX trap.

Figure 2a (b), shows the higher (lower) power confocal PL electric field map of the free and trapped IX, showing the difference in dipole energy shift for the two peaks. Scans are taken with different excitation power and collection time for ease of data display. The field is calculated by $E_{hs} = \frac{V_{tg}-V_{bg}}{t_{total}} * \left(\frac{\varepsilon_{hBN}}{\varepsilon_{TMD}}\right)$, where $E_{hs}$ is the electric field, $V_{tg}$ ($V_{bg}$) is the voltage applied to the top (back) gate, $t_{total}$ is the combined thicknesses of the encapsulating hBN layers, and $\varepsilon_{TMD}$ ($\varepsilon_{hBN}$) is the average of the vertical components of the relative permittivity of the TMD (hBN) layers.[16,17] The ratio of applied voltage to the top and back gate is determined by the ratio of hBN thicknesses $\alpha = 3.6$, such that the heterostructure remains charge neutral.[16] With no applied external field, the free IX peak appears with an energy of 1.35 eV. As the voltages to create the trapping potential are turned on a signal from trapped IXs appears at lower energy than free IXs. As the trapping voltage is increased, the energy of both IXs increases, and the PL intensity from the trapped IX signal increases. We fit a Lorentzian to the free and trapped IX peaks to extract their center energy for each electric field (Figure 2c). The peak amplitudes and linewidths from these fits are shown in Figure S6. The increasing electric field increases the free IX energy by a value of 0.59

eV/(V/nm), corresponding to an electron-hole separation of 0.59 nm for the free IXs. We measured the temperature dependence of the PL up to 20 K which showed stable trapped IX emission (Figure S7a). The trapped IX peak exhibited co-circularly polarized PL when excited with a 1.71 eV laser, on the WSe$_2$ neutral exciton resonance (Figure S7b), indicating that this is a spin-triplet IX associated with an H-type heterostructure.[13]

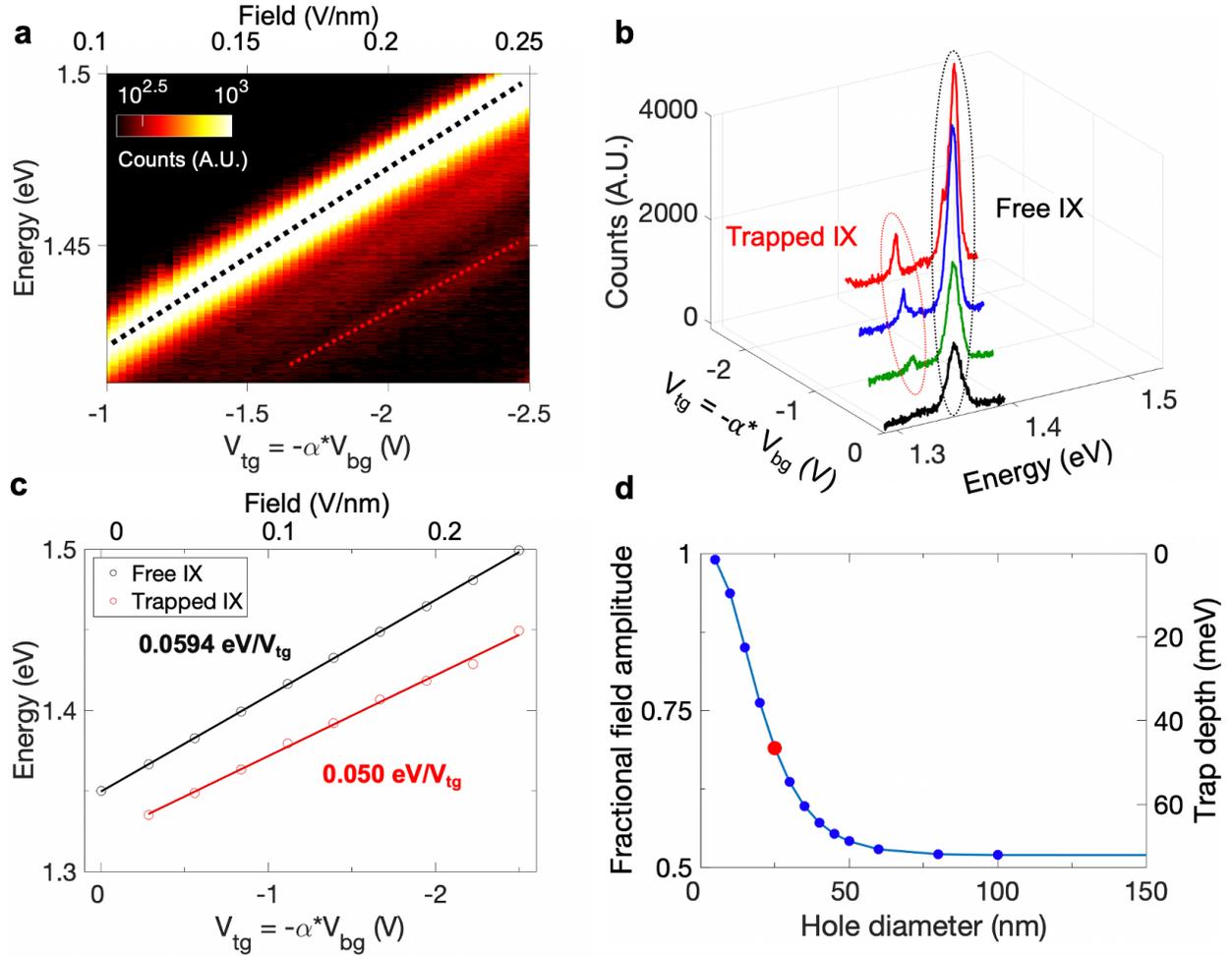

**Figure 2.** Confocal PL of the trapped and free signal. (a) High excitation power (10 µW) confocal PL as applied gate voltage is swept. The color bar is scaled to show trapped IX signal more prominently, and field on top axis represents the field magnitude far away from the hole. Black (red) line traces free (trapped) IX PL signal. (b) Low power (250 nW) excitation shows the free and trapped signals for several different trapping gate strengths. Black line shows PL signal with no applied field, green, blue, and red lines show PL signal as fields of increasing magnitude are applied. (c) Peak center positions of the free and trapped IX at 250 nW excitation as a function of gate. The field on the upper axis of the plot corresponds to the electric field magnitude away from the hole. (d) COMSOL simulation data of the fractional electric field amplitude for several hole diameters. The right y-axis represents the trap depth for the IX at maximum applied voltage. Red dot represents the data point closest to the measured trapping potential of 48 meV.

Figure 2c shows that the trapped IX peak has a dipole shift that is 84±2% that of the free IX, indicating that the field is 84% as strong under the hole. Figure 2d shows COMSOL modelling of the electric field profile as a function of hole diameter. The fractional electric field amplitude is the electric field magnitude in in the center of the hole at the height of the TMD heterostructure divided by the electric field magnitude far from the hole, for the given thicknesses of hBN encapsulating layers. This fractional field strength should be the ratio of the dipole shifts of the trapped and free IXs. From this model, we extract a hole diameter of ~15-20 nm based on the trapped to free dipole shift ratio of 84%. Furthermore, the model gives a depth of the IX trapping potential for a given applied voltage, shown on the right axis of Figure 2d for the strongest voltage applied (0.25 V/nm). For a ~20 nm hole, the corresponding trapping energy is ~40 meV, which is consistent with our measured value of 48 meV. This dipole shift ratio is our primary evidence for the nanoscale nature of the trapping. Figure S5 shows results from a device with a 300 nm hole, for which the fractional dipole shift reaches the limiting case for large diameter holes of 50%.

Figure 2c also shows a difference in the zero-crossing of the trapped and free IX energies, which we would expect to intersect at zero applied field. We attribute this offset to either a small built-in electric field, or to a change in the IX energy due to the change of dielectric environment (i.e., the absence of graphene above the trap).[24,25]

We note that at low excitation power (250 nW), the spectrally integrated PL from trapped IX can be as strong as 15% of the free IX (see Figure S8), despite the fact that the trap occupies less than 1% of the confocal PL collection area. The relatively large intensity of the trapped signal could be the result of several mechanisms. First, the trapped IX has a lower energy, so the free IX can relax to the trapped IX before its decay. Second, the non-radiative decay rate could be suppressed for the trapped IX due to the suppressing of IX diffusion.[26] Additionally, we note that when a field is applied in the opposite direction, such that the IX energy is decreased, there is no additional signal from the removed graphene area (see Figure S9). This indicates that there are no "anti-trapped" excitons, consistent with previous reports on exciton diffusion.[17]

We performed power dependent PL measurements to probe the density dependence of IXs. Figure 3a,b show the IX spectra of both the trapped and free IX for various excitation powers, showing that the trapped IX signal saturates at a significantly lower power than the free IX signal, which is supporting evidence towards the nanoscale nature of the trapping potential. The trapped and free IX are fit with a sum of two Lorentzians, and their amplitudes are plotted as circles in Figure 3c as a function of excitation power. The solid lines in Figure 3c are fits to the equation $I = \beta \frac{P}{1+P/P_{sat}}$ where I is the amplitude of the fitted PL intensity, P is the excitation laser power, and $\beta$ and $P_{sat}$ are fit parameters representing the fit constant and saturation power. Fitting to peak amplitudes in Figure 3c yields a saturation power of 330 nW for the trapped IX, and 4 µW for the free IX.

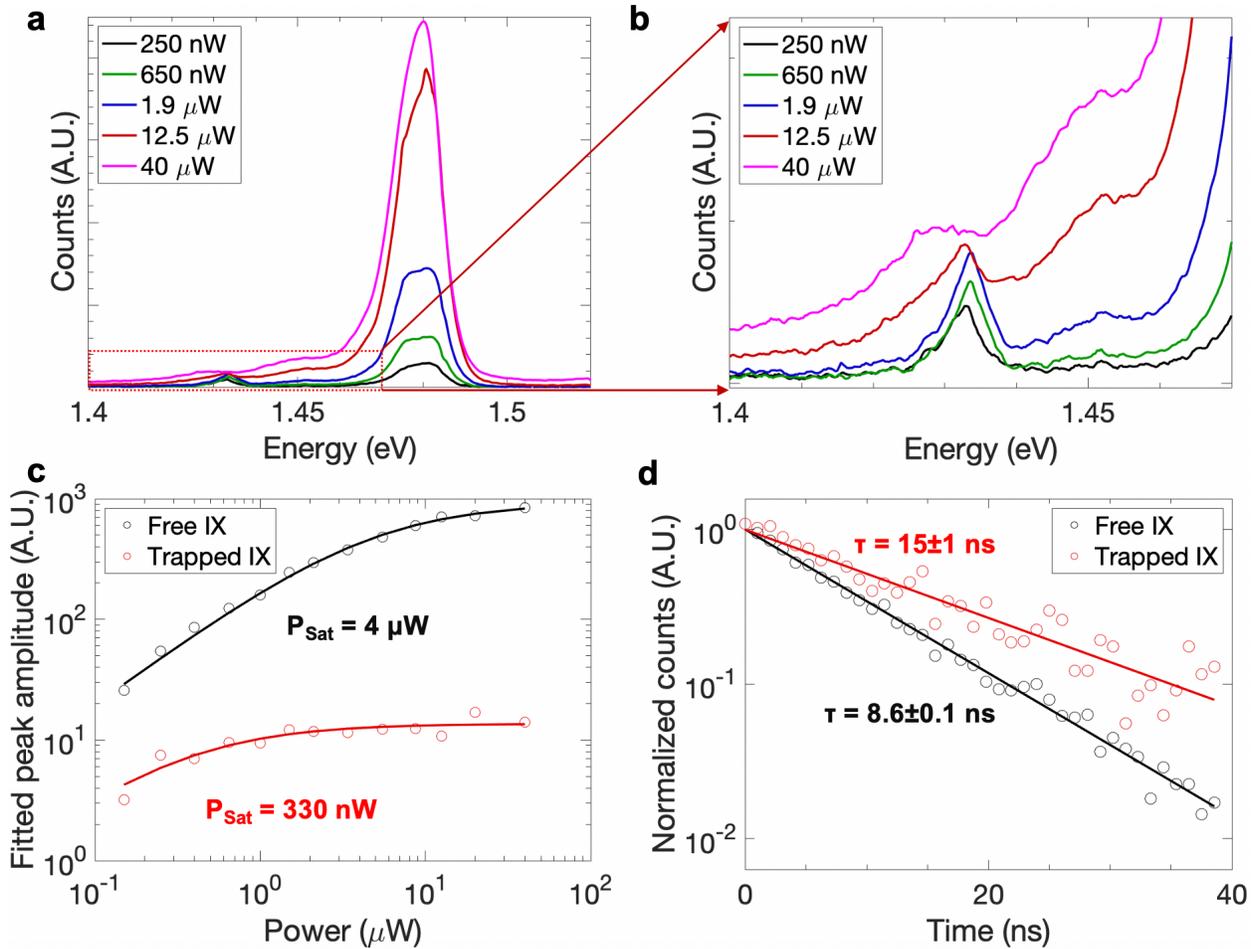

**Figure 3.** Power and temporal dependence of the trapped and free signals. (a) Confocal PL signal from the trapped and free signal at near maximum applied voltage (0.22 V/nm field away from the hole). (b) Zoom in on the trapped signal from (a). (c) Fitted signal intensity as a function of power showing low power (330 nW) saturation of the trapped IX. Dots show data points, with line as fit to the saturation equation discussed in text. (d) Time dependent PL of the trapped and free signal measured at an electric field of 0.2 V/nm away from the hole. Trapped and free IX signals were fit with single exponential decay functions, giving lifetimes of 15 ns and 8.6 ns respectively.

In order to probe the dynamics of IXs in the nanoscale trap, we performed time dependent PL spectroscopy using pulsed excitation with a 712 nm laser and a time correlated single photon counting setup, with an applied free field of 0.2 V/nm. The time resolved PL of the trapped and free IX are shown in Figure 3d, which reveal that the trapped IX has a longer lifetime (15 ns) than the free IX (8.6 ns), with full data including the laser pulse profile in Figure S10. The longer lifetime of the trapped IX is again consistent with the suppression of IX diffusion effects and weaker electric field experienced within the trap.[16] When the free electric field is lowered to 0.15 V/nm, slightly less than the expected field experienced by the trapped exciton in Figure 3d, the lifetime of the free IX reduces to 11 ns, showing that the weaker electric field cannot account for the full difference of lifetime between the trapped and free IXs (see Figure S11).

**Conclusion and Discussion:**

In summary, we have demonstrated deterministic trapping of IXs on a nanoscale level via a nanopatterned graphene gate. Unlike previous studies which relied on nanopillars,[27] defects, or moiré potentials to trap excitons in 2D materials, here, we show deterministically placed (lithographically defined) confinement of an excitonic system which allows for an extremely broad electronic tuning of the emission energy of ~100 meV. We note that previous studies that have investigated spatial control of IX were performed on 1 μm slot structures that were not able to realize the nanometer scale quasi-zero dimensional confinement shown here.[17] Additionally, there have been previous reports of quantum dot like behavior in various 2D materials, either from defects in the crystal lattice or by unique electrostatic gating of monolayer TMDs,[28–31] but none of these works have demonstrated stable trapping of optically active IXs. Our approach opens up a new avenue to realize deterministically placed IX quantum dots that make use of the highly tunable energy of the IX energy. In future studies, we plan to refine our architecture decreasing the trap diameter and increasing the strength of the confinement potential through ion beam lithography and dielectric engineering approaches[32] to ultimately trap single stable IXs which can be integrated with on-chip photonic or plasmonic structures for scalable quantum technology applications.

**Methods:**

**Device fabrication**: All 2D layers were mechanically exfoliated from bulk crystals onto 285 nm $SiO_2$/Si wafers and identified by optical microscopy and characterized with atomic force microscopy. The TMD monolayer crystal axes were determined by polarization resolved second harmonic generation measurements, using an 800 nm, 120 fs pulsed laser.[33,34] Layers were aligned by optical microscope, with TMD monolayers aligned to ~5.5° relative rotation of their crystal axes, and transferred on top of one another using a dry transfer technique, using a polycarbonate film on polydimethylsiloxane.[35] We note that the relative twist angle between the TMD layers is large (~5.5° away from 60°) so that the moiré wavelength (~3.5 nm) is small relative to the gate defined trapping potential. The full device structure from top to bottom is bilayer graphene/6 nm hBN/ML $MoSe_2$/ML $WSe_2$/20 nm hBN/few layer graphene.

The top FLG was nano-patterned via by electron beam lithography (EBL) and reactive ion etching (RIE). A thin layer of 950 poly(methylmethacrylate) (PMMA) A2 was spun at 3000 rpm to produce a roughly 80 nm thick film on top of the stacked device. EBL was performed in an Elionix Model ELS-7000, with an electron beam current of 50 pA and an aperture width of 60 μm. PMMA was developed in a 1:4 MIBK:IPA solution for 30 seconds, with light stirring. The sample was then placed in an inductively couple plasma reactive ion etch (Plasmatherm Versaline DSE III), with 40 sccm $O_2$ gas at a pressure of 100 mTorr for 3 seconds, in order to etch the exposed area of graphene. The PMMA was washed away in acetone and the sample was cleaned in isopropyl alcohol, and contact mode AFM squeegee.[36] We then performed a second round of EBL in order to evaporate Cr/Au contacts to electrically connect to the graphene gates and TMD layers.

**Optical measurements**: All PL spectroscopy was in an optical cryostat at 4 K. The sample was excited with a 670 nm diode laser except for the time and polarization resolved measurements which were performed with a 712 nm laser. Gate voltages were applied to the graphene top and bottom gates, while the TMD layers were connected to virtual ground. The confocal PL measurements were performed with a 0.6 NA 40x objective and a confocal pinhole which results in a collection diameter of 1.5 μm on the sample. The PL signal was measured with a grating based spectrometer and cooled CCD camera. In the polarization resolved measurements, combinations of polarizers, half wave plates and quarter wave plates are used. The time resolved measurements were performed via time resolved single photon counting with a silicon avalanche detector and picosecond event timer. In the time resolved measurements, we used an acousto-optic modulator (AOM) to generate 20 ns square pulses from a continuous wave laser. The time resolved PL (Figure 3d) are plotted after the square pulse is less than 5% its maximum intensity.

**COMSOL simulations**: All COMSOL models used 2D axisymmetric modelling to represent a rotationally symmetric hole, using known values for the in plane and out of plane dielectric constants of hBN and TMD layers, and hBN thicknesses that match AFM data of the individual 2D layers. Voltages applied to top and back gates were set at the graphene-hBN interfaces, while the top and bottom interfaces of the TMD heterostructure were grounded. Electric field data were calculated by averaging the field above and below the TMD heterostructure layer.

**Supporting Information**: Device images, AFM topography, additional PL and lifetime data and analysis.

**Acknowledgments:**

**General:** The authors acknowledge useful discussions with Andrea Young and Lazaro Calderin.

**Funding:** This work is mainly supported by the National Science Foundation (Grant Nos. DMR-1708406, DMR-1838378, and DMR-2054572). DGM acknowledges support from the Gordon and Betty Moore Foundation's EPiQS Initiative, Grant GBMF9069. K.W. and T.T. acknowledge support from the Elemental Strategy Initiative conducted by the MEXT, Japan, Grant No. JPMXP0112101001, JSPS KAKENHI Grant No. JP20H00354 and the CREST(JPMJCR15F3), JST. HY acknowledges support from Guangdong project No. 2019QN01X061. JRS acknowledges support from AFOSR (Grant Nos. FA9550-17-1-0215, FA9550-18-1-0390, and FA9550-20-1-0217) and the National Science Foundation Grant. No. ECCS-1708562. BJL acknowledges

support from the Army Research Office under Grant nos. W911NF-18-1-0420 and W911NF-20-1-0215. Plasma etching was performed using a Plasmatherm reactive ion etcher acquired through an NSF MRI grant, award no. ECCS-1725571.

**Author Contributions:**

JRS, BJL and DNS conceived the project. JRS and BJL supervised the project. DNS modelled and fabricated the structures, and performed the experiments, assisted by FM, CM and AA. DNS analyzed the data with input from JRS and BJL. MRK and DGM provided and characterized the bulk $MoSe_2$ and $WSe_2$ crystals. TT and KW provided hBN crystals. HY provided theoretical support in interpreting the results. DNS, JRS, BJL, and HY wrote the paper. All authors discussed the results.

**Competing Interests:**

The authors declare no competing interests.

# Supporting Information for

# Nanoscale trapping of interlayer excitons in a 2D semiconductor heterostructure

**Author Names:** Daniel N. Shanks[1], Fateme Mahdikhanysarvejahany[1], Christine Muccianti[1], Adam Alfrey[1], Michael R. Koehler[2], David G. Mandrus[3-5], Takashi Taniguchi[6], Kenji Watanabe[7], Hongyi Yu[8], Brian J. LeRoy[1], and John R. Schaibley[1]*

**Author Addresses:**
[1]Department of Physics, University of Arizona, Tucson, Arizona 85721, USA
[2]JIAM Diffraction Facility, Joint Institute for Advanced Materials, University of Tennessee, Knoxville, TN 37920
[3]Department of Materials Science and Engineering, University of Tennessee, Knoxville, Tennessee 37996, USA
[4]Materials Science and Technology Division, Oak Ridge National Laboratory, Oak Ridge, Tennessee 37831, USA
[5]Department of Physics and Astronomy, University of Tennessee, Knoxville, Tennessee 37996, USA
[6]International Center for Materials Nanoarchitectonics, National Institute for Materials Science, 1-1 Namiki, Tsukuba 305-0044, Japan
[7]Research Center for Functional Materials, National Institute for Materials Science, 1-1 Namiki, Tsukuba 305-0044, Japan
[8]Guangdong Provincial Key Laboratory of Quantum Metrology and Sensing & School of Physics and Astronomy, Sun Yat-Sen University (Zhuhai Campus), Zhuhai 519082, China

**Corresponding Author:** John Schaibley, johnschaibley@email.arizona.edu

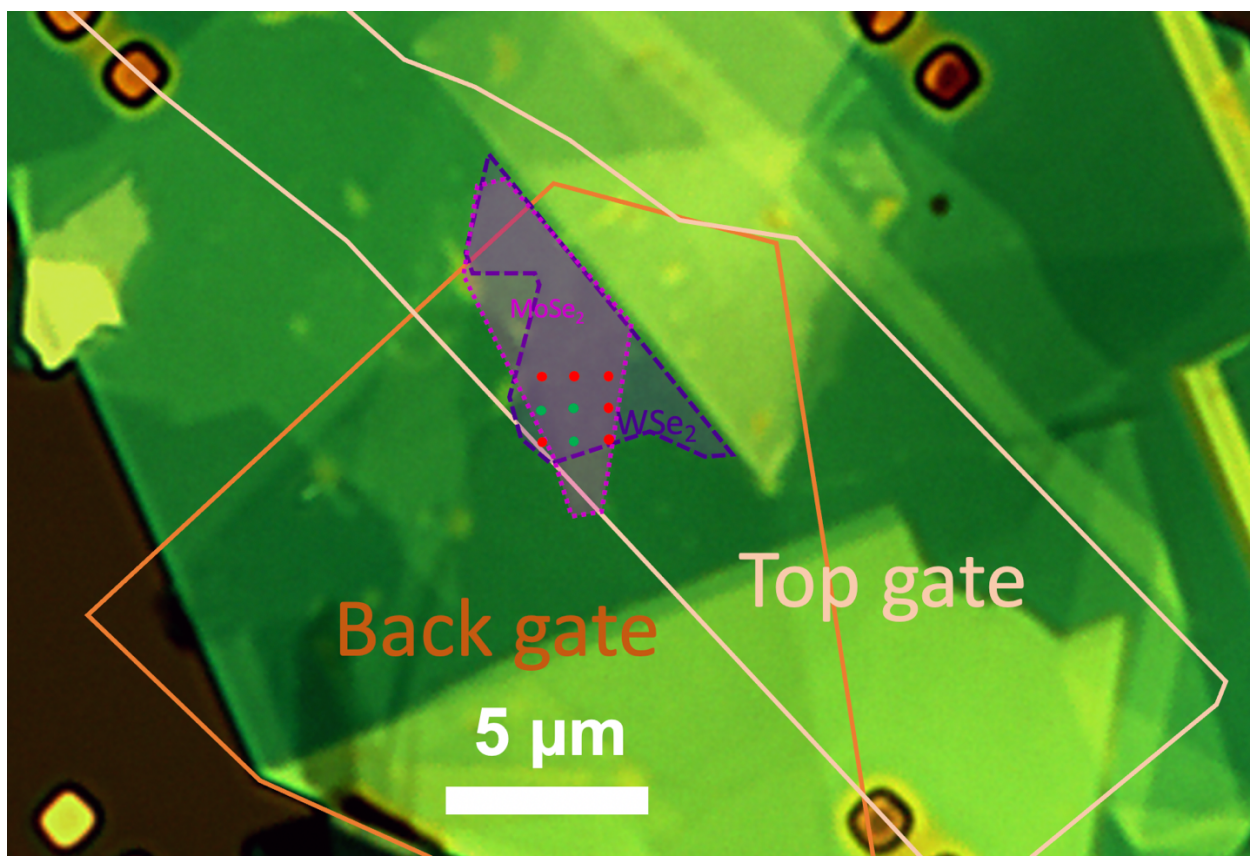

**Figure S1.** Optical microscope image of the device, with layers outlined. WSe$_2$ shown in purple, MoSe$_2$ shown in pink. Top and bottom graphene gates outlined. Green (red) dots show locations of holes which have (have not) shown PL emission from trapped excitons. TMDs are electrically connected to one another by attached bulk, which protrudes from the top hBN encapsulation layer to allow for electrical contact. We observe trapped IX PL from three of the nine hole sites. The lack of trapped IX PL from the other holes sites is attributed to contamination in the fabrication process or to their proximity to the sample edge.

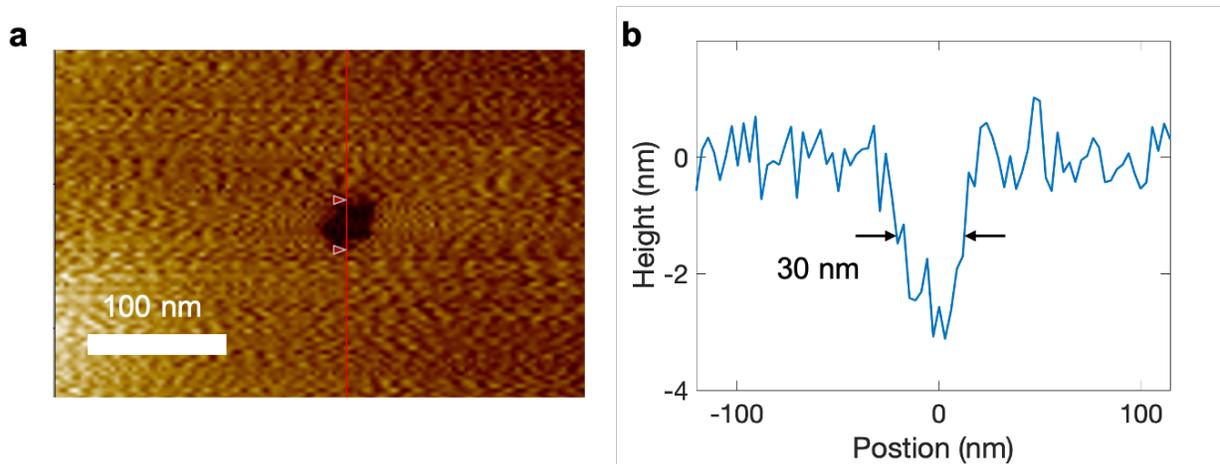

**Figure S2.** AFM Topography of the nanoscale graphene hole. (a) AFM topography image of the removed hole in few layer graphene after AFM contact mode squeegee cleaning. (b) Line cut of the AFM topography corresponding to the red line in (a), showing a hole diameter of ~30 nm.

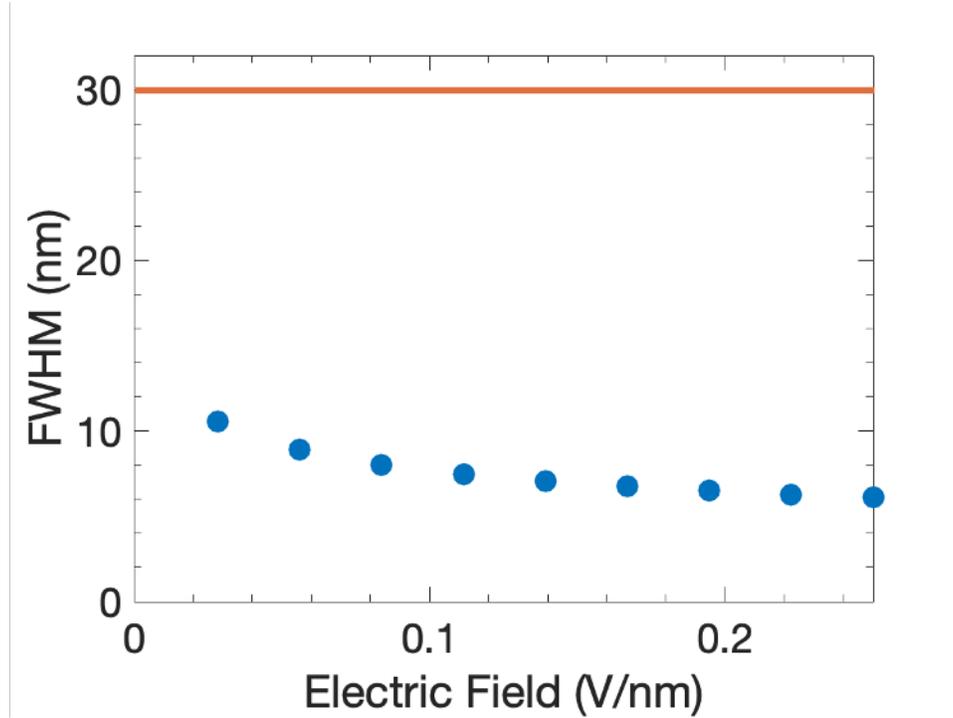

**Figure S3.** Characteristic IX envelope full width at half maximum as function of applied electric field, for the electric fields shown in Figure 2. The trap frequency, $\omega$, is found by a quadratic fit to the IX potential energy profile, and $m$ is taken to be sum of the $WSe_2$ effective hole mass[1] and the $MoSe_2$ effective electron mass[2], giving $m = 1.25\ m_0$. The orange line represents the graphene hole diameter.

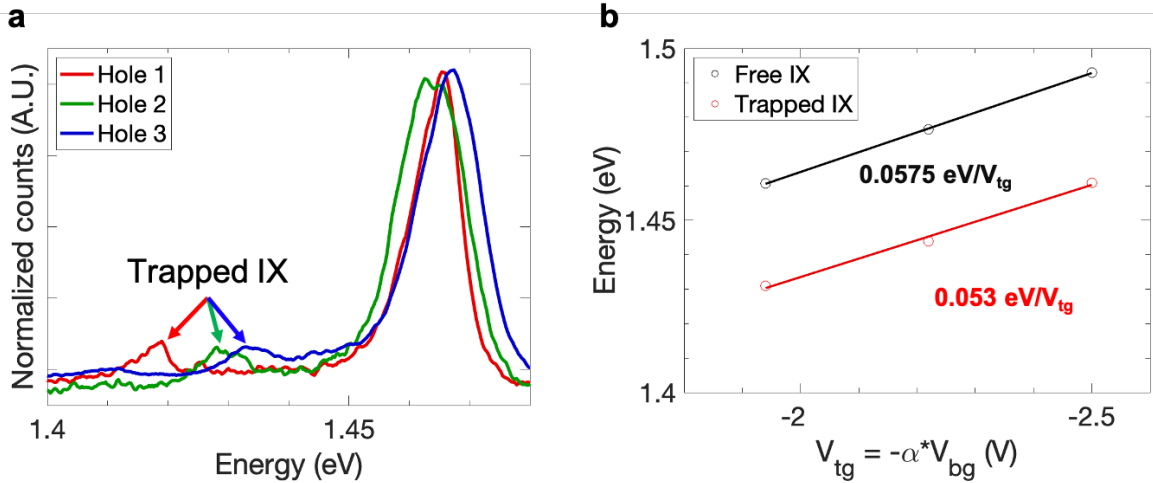

**Figure S4.** Photoluminescence from multiple hole locations on primary device. (a) PL spectra from the three different holes where a trapped IX signal was observed at 80% of maximum gate. Hole 1 corresponds to data in main text. (b) Free and Trapped PL energy as a function of applied gate for hole 2. On this hole, the trapped IX occurs at higher energy, with a smaller dipole shift compared to the PL data from hole 1. The trapped to free IX dipole shift ratio on hole 2 is 0.92, compared to 0.84 on hole 1, resulting in a hole diameter of ~10-15 nm. Variations in the energy and dipole energy shift of the PL can be attributed to varying hole size. The trapped IX signal was not observed from hole 2 for gate values lower than those shown in (b).

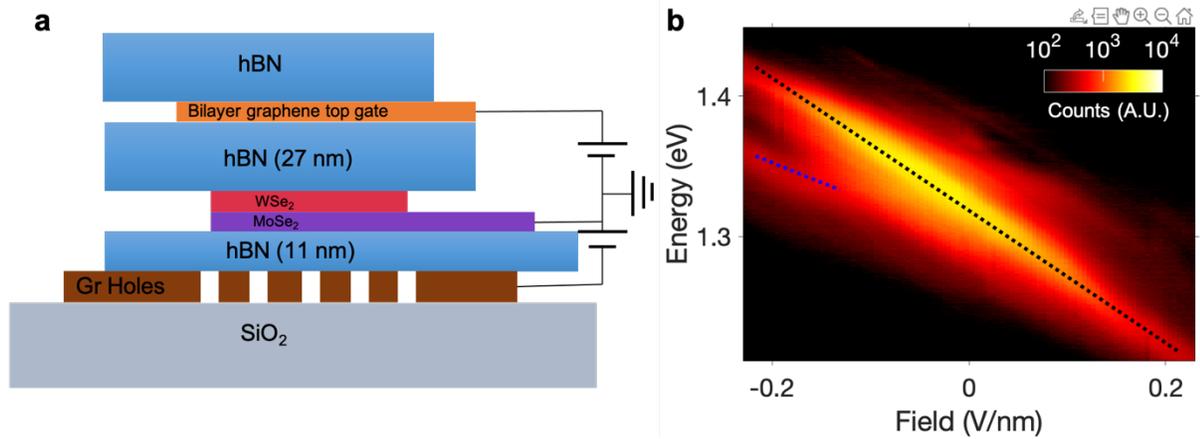

**Figure S5.** Layout and trapping of IXs in device B. (a) Layout of device B. Larger, 300 nm diameter graphene holes are etched first, and then TMD heterostructure is transferred onto holes. (b) PL spectra with gate sweep and high power excitation (15 μW), sweeping field from -0.23 V/nm to 0.23 V/nm. Dipole shift direction is flipped, as expected for a heterostructure with $WSe_2$ on top. Black (blue) line shows free (trapped) IX signal, with a 0.45 (0.23) eV/(V/nm) dipole shift.

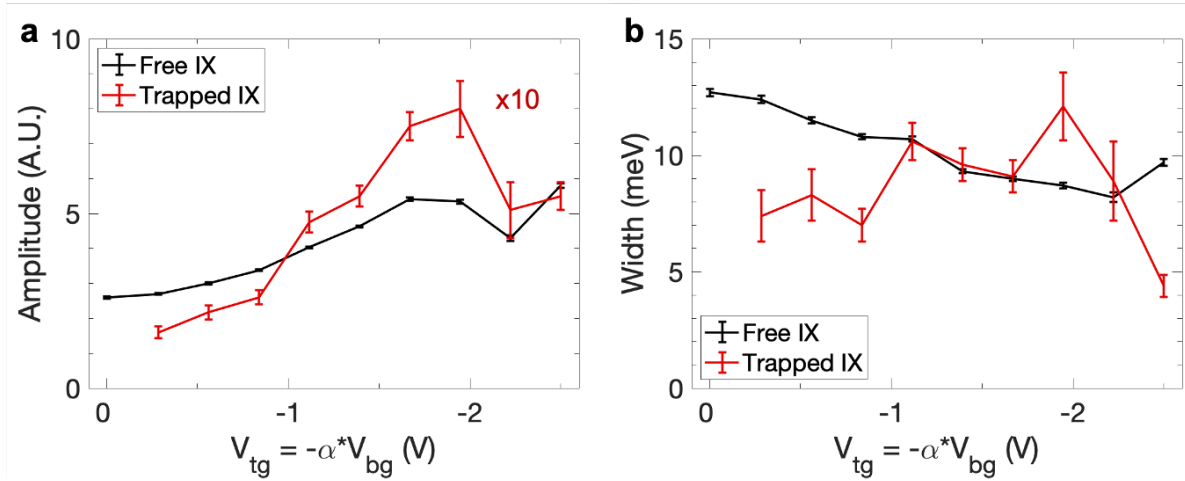

**Figure S6.** Amplitude (area under curve) (a) and width (full width half max) (b) of Lorentzian fits to free and trapped IX peaks in low power confocal PL scans, corresponding to data shown in Fig. 2b. Error bars correspond to the standard error in the Lorentzian fit.

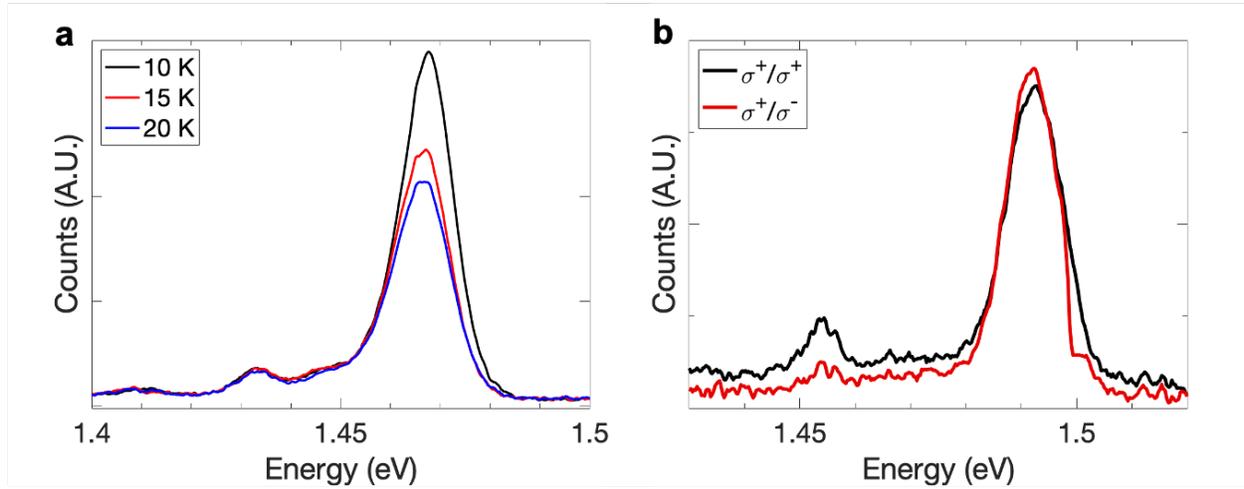

**Figure S7**. (a) Temperature dependent PL of the trapped and free signal, with 0.2 eV/(V/nm) applied. (b) Co (black) and Cross (red)-circular polarized PL of the trapped and free IX, when exciting with 1.71 eV at maximum trapping potential. The free IX signal is unpolarized, and the trapped signal retains co-circular polarization. We note that the difference in PL polarization could be attributed to the electric field as shown in [3].

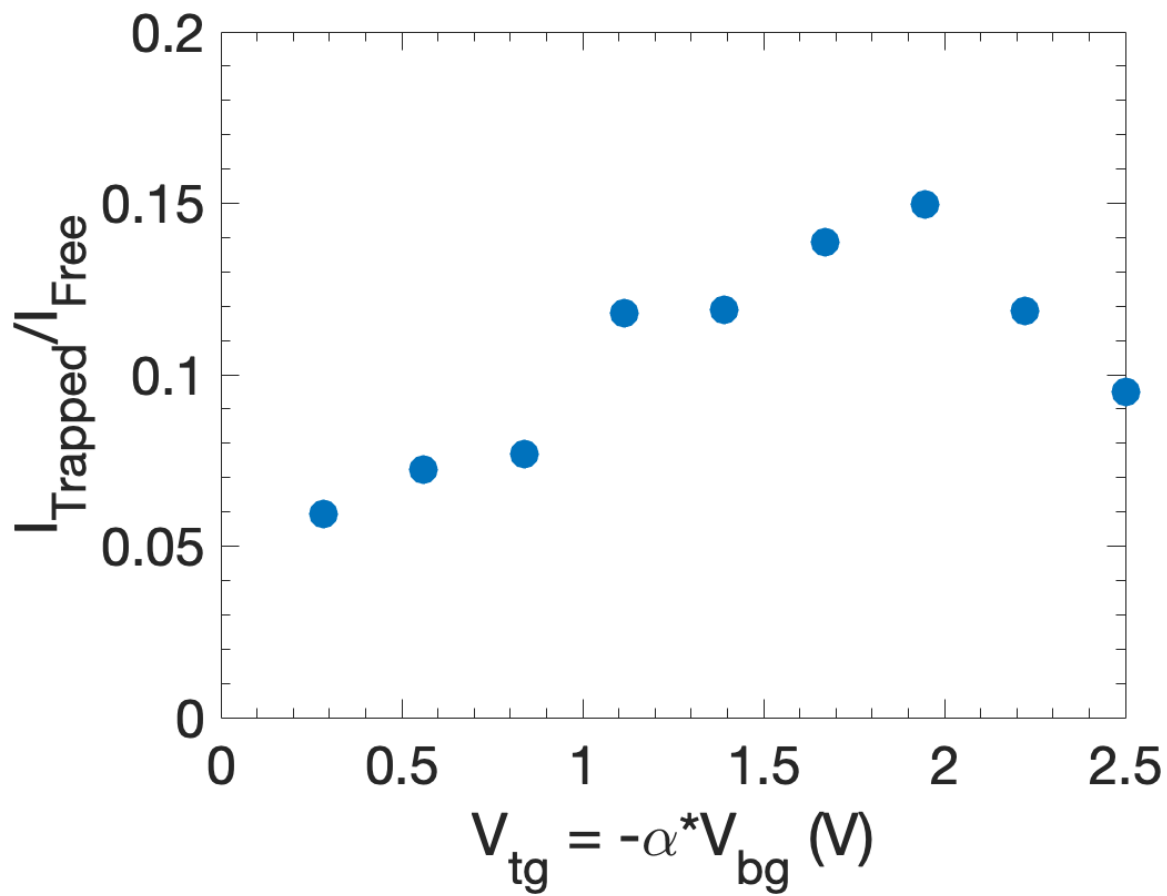

**Figure S8.** Ratio of trapped to free IX PL intensity as a function of gate for 250 nW excitation power, using data from Fig. 2b. The ratio is determined by fitting each spectrum to a sum of two Lorentzians and comparing the amplitude of the fits.

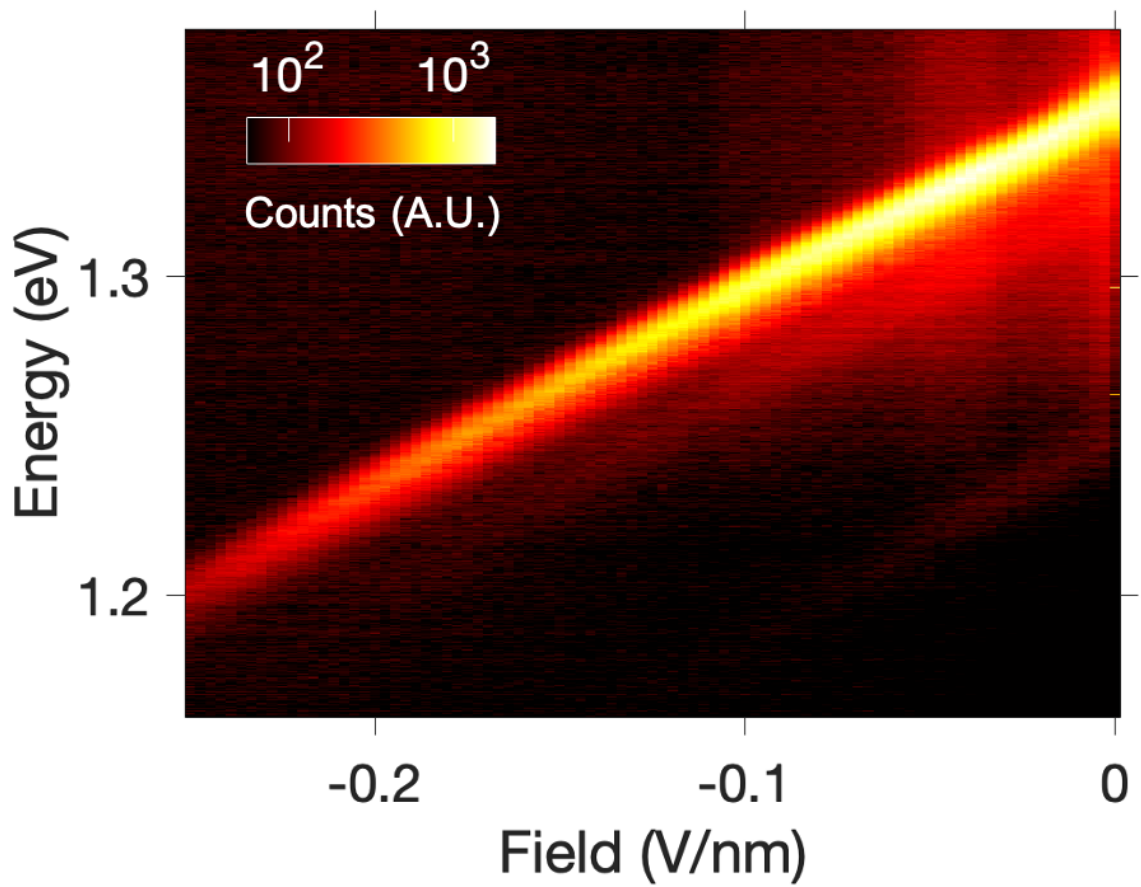

**Figure S9.** Anti-trapping electric field map. Confocal PL electric field map with the electric field pointing in the same direction as the dipole moment of the IX, with an "anti-trapping" configuration of the spatial IX energy. There is no signature of an anti-trapped IX, which would appear at higher energy than the free IX, while the voltage is set such that the exciton will be forced away from the hole. We believe the anomalous peak starting at 1.2 eV, -0.05 V/nm is from IX's bound to $WSe_2$ defects. This peak appears at lower energy than the free IX peak but lacks the defining characteristics of the trapped IX signal, namely a lack of spatial localization near the hole sites, disappearance at zero trapping field, and reduced dipole energy shift.

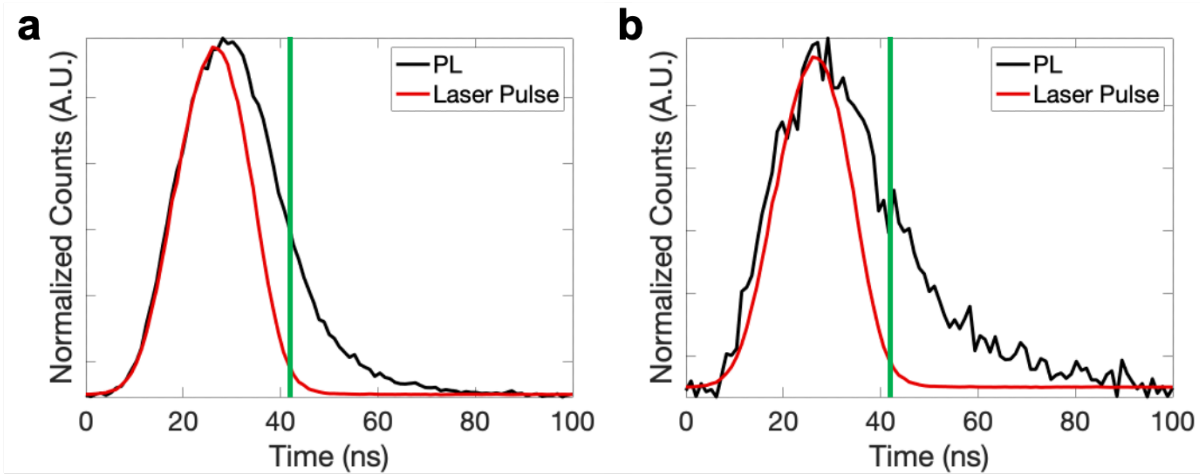

**Figure S10.** Time dependent PL data including laser pulse of the free (a) and trapped (b) signals, corresponding to data in Fig. 3d. In our experiment, we use a ~20 ns pulse with a fast fall time. We then cut off the data at the point where the laser pulse was less than 5% of its maximum intensity, and only fit to the data after the falling edge of the pulse (after the green line).

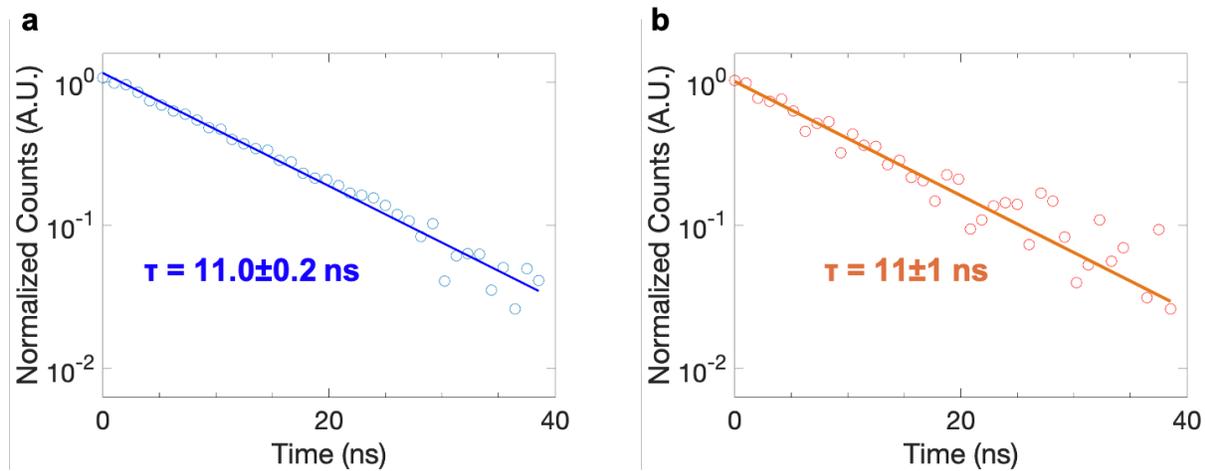

**Figure S11.** Lifetime of the free and trapped IX at varied gate voltages. (a) Lifetime measurement of the free IX, with the applied field decreased to 0.15 V/nm. (b) Lifetime measurement of the trapped exciton with the gates set to maximum, such that the free field is 0.25 V/nm and the field experienced by the trapped IX is ~0.2 V/nm. The lifetime in (b) still remains higher than the free IX lifetime for 0.2 V/nm of 8.6 ns, further supporting the conclusion that the trapping of the IX contributes to the increased lifetime.